# The combination of directional outputs and single-mode operation in circular microdisk with broken $\mathcal{PT}$ symmetry

Qinghai Song[1, 2, *], Jiankai Li[1], Wenzhao Sun[1], Nan Zhang[1], Shuai Liu[1], Meng Li[1], Shumin Xiao[3, †]

1. Integrated Nanoscience Lab, Department of Electrical and Information Engineering, Harbin Institute of Technology, Shenzhen, 518055, China
2. State Key Laboratory on Tunable Laser Technology, Harbin Institute of Technology, Harbin, 150080, China
3. Integrated Nanoscience Lab, Department of Material Science and Engineering, Harbin Institute of Technology, Shenzhen, 518055, China

\* Qinghai.song@hitsz.edu.cn
† shuminxiao@gmail.com

## Abstract:

Monochromaticity and directionality are two key characteristics of lasers. However, the combination of directional emission and single-mode operation is quite challenging, especially for the on-chip devices. Here we propose a microdisk laser with single-mode operation and directional emissions by exploiting the recent developments associated with parity-time (PT) symmetry. This is accomplished by introducing one-dimensional periodic gain and loss into a circular microdisk, which induces a coupling between whispering gallery modes with different radial numbers. The lowest threshold mode is selected at the positions with least initial wavelength difference. And the directional emissions are formed by the introduction of additional grating vectors by the periodic distribution of gain and loss regions. We believe this research will impact the practical applications of on-chip microdisk lasers.

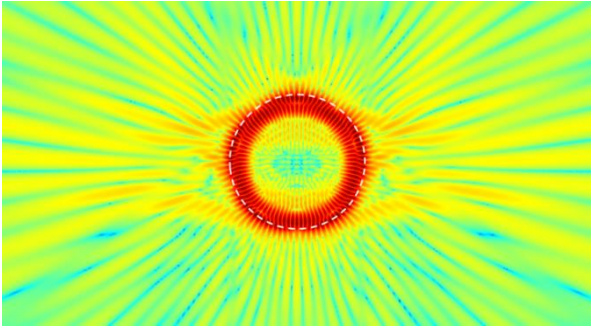

*Logscale plot of field pattern and directional emissions of whispering gallery modes in circular cavity with PT symmetry breaking*

# Introduction

Parity-time (PT) symmetric systems have attracted considerable interest due to their unique properties, e.g. unidirectional reflectionless, coherent perfect absorber, non-reciprocity, and loss-induced threshold reduction. [1-10] Very recently, the researches on PT symmetry have been extended from parallel balanced gain-loss waveguide to optical microcavities, in which gain and loss are equally applied in a manner that $n(-x) = n^*(x)$. Several types of novel lasing behaviors have been successfully proposed and experimentally observed. Based on the finding that both pole and *zero* of S matrix reach the real axis, Longhi [11], Chong [12], and Ge [13] have independently proposed several types of PT symmetric laser absorber, which has been recently experimentally verified by and Sun et al [14] and Gu et al [15]. By exploiting the periodic modulation along the Azimuthal direction ($n(-\phi) = n^*(\phi)$), Ge and Stone have proposed the thresholdless PT symmetry breaking [16], which has been utilized by Feng et al. in microring to achieve single-mode laser emission [17].

Directionality is another essential characteristic of on-chip laser. Due to the presence of rotational symmetry[18, 19], the conventional microdisk laser usually lack of directional output. This shortage has been overcome by tailoring the boundary of microdisk laser, which is pioneered by Nöckel and Stone [20]. In past decades, many types of directional outputs have been proposed and experimentally observed in microdisks, e.g. directional laser emissions in quadruple microdisk [21, 22], unidirectional laser emission in spiral-shaped microdisk [23-25], limacon microdisk [26-29], and waveguide connected microdisks [30, 31] et al. The conventional studies on deformed microcavities rely on the tradeoff between cavity Q factor and directionality. Consequently, the Q factors of above proposals are usually orders of magnitude lower than a circular microdisk or microring. Recently, by applying the perturbation to modify the components of angular momentum, a combination of unidirectional output and Q factor around half of a circular cavity has been obtained simultaneously [32]. Up to now, both the PT symmetric laser and directional laser emission have been substantially developed in optical microcavities. However, the combination of directional output and PT symmetric laser with single-mode operation is still absent. Here, by exploiting the periodically distributed gain and loss, we explore the possibility to obtain single-mode laser and directional output simultaneously.

## Results and Discussions

### Single-mode operation in PT symmetric microdisk

The cavity in this research is the typical two-dimensional circular cavity, which has been widely studied before. It is well known that the resonances inside the circular cavities are the whispering gallery modes that are confined along the cavity boundary. In principle, circular cavity can support neither single-mode operation nor the directional output. Here we consider its corresponding behaviors under PT symmetry. The distributions of gain and loss are schematically depicted in the inset in Fig. 1(b). They are one-dimensional periodic stripes of gain and loss with equal width $\Lambda/2$. The refractive indices of two regions are set as $n_{g(l)} = n_0 \pm n''i$. At first glance, these resonances shall remain neutral due to the balanced gain and loss. However, this situation is changed once the mode coupling is considered. In a simple two-mode interaction model, the interaction between two modes usually generates two hybrid modes with modified field distributions. One hybrid mode that is possible to be mainly confined within the gain area and thus reach the lasing threshold easily. On the contrary, the other mode is mostly confined in the absorptive region and becomes lossier. These two hybrid modes are usually known as lasing and absorptive modes in the recent literatures.

Based on the qualitative analysis, we then numerically studied the PT symmetry inside circular cavity. The cavity is considered as a two-dimensional object by applying the effective refractive index and studied with finite element method (Comsol multiphysics 3.5a). Here the radius of cavity R = 5 μm and the periodicity Λ = 1 μm. The refractive index $n_0$ is 1.56 for polymer or photoresist. The outgoing waves were absorbed by a perfect matched layer at far field, leading to quasibound states with complex eigenfrequencies (ω). The resonant wavelengths are defined as λ = 2πc/ω and a factor β ($\beta = Im(\omega)/100c$, here *c* is the speed of light in vacuum and the unit is cm$^{-1}$) is defined to estimate the gain or loss. In our simulation, only TE polarization (E is in plane) is considered. By setting n" = 0.009, we have checked a wide range of wavelength from 800 nm to 1100 nm to study the lasing actions inside it. As what we analyzed above, most of the whispering gallery modes are either neutral or strongly lossy due to the balanced gain and loss. Only a few resonances reach thresholds and have positive β values. And the β factor at 919.8 nm is much larger than all the others.

To understand the formation of lasing actions around 919.8nm, we have studied the corresponding lasing behaviors by varying n". All the results are summarized in Figs. 1(a) and 1(b). With the increase of n", we can see that two whispering gallery modes approach each other. The mode separation distance is around 0.2 nm at the beginning and quickly reduces to ~ 0.015nm at n" = 0.0025. After that, their resonant wavelengths change very slowly and two modes cross at n" ~ 0.007. Meanwhile, the β values show different behaviors. Two resonances remain at very small negative values at the beginning. Once n" is above 0.0017, two resonances depart to a pair of inversed moduli. One mode is lasing mode with positive β value. The other mode is absorptive mode with negative β. The approaching in wavelength and bifurcation in β are both consistent with the conventional PT symmetric behaviors in literatures and can be easily explained by the mode coupling model [4].

Figure 1(c) shows the field patterns of resonances marked 1-6 in Fig. 1(a). At n" = 0, two resonances are regular whispering gallery modes with mode numbers (*m = 47, l = 1*) and (*m = 42, l = 2*). Here *m* and *l* are the Azimuthal number and radial number, respectively. Without external modulation, these two modes are two independent resonances. Once the gain and loss are added (n" is increased), these two modes start to interact with each other due to the overlapping in both real space and frequency spectra. Similar to the conventional studies of mode coupling in passive cavities, the interference between fields of two resonances can significantly change their corresponding field distributions. We can clearly see the field distribution of mode-3 is mainly confined within the gain region, whereas mode-4 is mostly localized within the absorptive region. Therefore, further increasing n" can both dramatically improve the amplification of one mode and enhance the absorption of the other mode. Consequently, two modes bifurcate at n" = 0.017. One mode is lasing mode with positive β and the other one is absorptive mode with negative β.

Besides the mode at 919.8nm, there are also some other modes have positive β values. We then calculate the dependence of resonant wavelength and β values on n" to explore the corresponding laser behaviors. All the results are summarized in Figs. 2(a) and 2(b), in which the second highest mode around 938.6 nm has been analyzed. We can see both the resonant wavelength and the β value are very similar to the results in Figs. 1(a) and 1(b) except the bifurcation point. In Figs. 2(a) and 2(b), their imaginary parts bifurcate at n" ~ 0.005, which is about 3 times larger than that in Fig. 1. As the β values of two modes remain around unit for a wider range of n", the amplification of lasing mode with fixed n" in Fig. 2(b) is thus smaller than that of mode in Fig. 1(b).

From the typical mode coupling theory, the bifurcation point is dependent on the coupling constant [2, 10, 17]. Then larger bifurcation point indicates that strong coupling constant is required to enable their mode interaction. In this sense, the wavelength difference between two initial modes in passive cavity should be larger. This is exactly what we can see in Fig. 2, where the wavelength detuning between two resonances at n" = 0 is around 0.55 nm, which is much larger than that in Fig. 1. Figure 2(c) shows the relation between the position of bifurcating point and the wavelength difference. We can see that the trends of wavelength difference in passive cavity matches the change of bifurcation in β, consistent with the coupled mode theory well.

In circular cavity, the whispering gallery modes with *l = 1* and modes with *l = 2* have different mode spacing. Consequently, they shall gradually approach each other and cross at some particular wavelengths. Considering the results in Fig. 2(c), it is easy to imagine that the mode with least wavelength difference shall bifurcate first in their imaginary parts of frequencies and thus have the largest β value. Then the wavelength difference between two nearby lasing modes will be significantly extended, which is similar to the well-known Vernier effect [33, 34] and inversed Vernier effect [35, 36]. This can be observed in Fig. 2(d), where n" is set at 0.005 for an example. It is easy to conclude that single mode laser can be achieved under such a pumping configuration.

**Directional outputs of PT symmetric laser**

In additional to single mode operation, it is also interesting to explore the directionality of the laser in circular cavity. The conventional whispering gallery modes in circular cavity are confined by total internal reflection. Their leakages are dominated by the tunneling along the tangential lines. Due to the rotational symmetry of circular cavity, the emissions of resonances are usually isotropic in plane (see Fig. 3(a) for an example). For the case of introducing balanced gain and loss to the cavity, the far field patterns will be totally different. Such a difference actually can be expected. As we know, a periodic modulation on gain and loss can also function as a grating[37]. Then the wave-vectors of resonances inside cavity will be changed by the additional vector of grating. In our design, as the grating is distributed in vertical direction, the wave-vector in this direction will be reduced and directional emissions in horizontal directions shall be formed.

Figure 3(b) shows the far field pattern of one lasing mode at 919.8 nm with n" = 0.0017. Different from the nearly isotropic emission in Fig. 3(a), here multiple directional outputs can be observed in the far field. This is consistent with above analysis. With the increase of n", the diffraction effect will be more dramatic due to the stronger modulation in refractive index. Thus the directional emissions in horizontal directions shall be improved. One example far field pattern of lasing mode with n" = 0.009 is shown in Fig. 3(c), where a "BAT" shaped far field distribution can be observed. Compared with Fig. 3(b), here the emissions along vertical directions have been further reduced. And the far field pattern is dominated by two "wings" along $\phi_{FF}$ = 0 and 180 degree. The divergence angles at full width half maximum are around 34 degree. Till now, we can say that the combination of single-mode laser operation and directional outputs have been achieved in a circular PT symmetric microdisk. By defining a character U as $U = \left(\int_{150}^{200} I(\phi_{FF})d\phi_{FF} + \int_{-30}^{20} I(\phi_{FF})d\phi_{FF}\right)/\int\int_{-150}^{200} I(\phi_{FF})d\phi_{FF}$ to record the percentage of energy within two main beams, the dependence of far field patterns on n" can be analyzed in detail [38]. As shown in Fig. 3(d), the U factor increases quickly at the beginning and then tends to saturation when n" is larger than 0.007. This shows that the far field patterns are improved by the stronger diffractive effect and consistent with the qualitative analysis well.

It is worth to note that utilizing grating to improve the directionality has been widely explored in microdisk lasers [39, 40]. However, the previous studies are mainly focused on the grating with modulation on $n_0$. Here the modulation is applied in the imaginary part of refractive index [37] and thus can easily combine with the research of PT symmetry breaking. In practical applications, one-dimensional modulation is not strongly dependent on high-cost equipments such as E-beam lithography. They can be easily realized by applying two-beam interference [37, 41]. The period Λ can be controlled by changing the intersection angle of two laser beams. Moreover, the period of gain and loss might not be well defined at Λ/2 = 500nm. We have also varied the Λ and studied the corresponding laser behaviors. Similar directional outputs and single-mode operation have been observed in a tested region 475nm < Λ/2 < 525 nm. This information makes our research to be more robust in fabrication. The period Λ is also dramatically changed to 4000 nm, directional single-mode laser emission has also been observed (see Figs. 4(a) and 4(b)). Due to the change of grating period Λ, the wave vectors along vertical direction are less cancelled and thus the far field patterns are quite different from Fig. 3(c). This is also consistent with the qualitative analysis well. It is worth to note that the cavity shape is not restricted to a circle. This research can also be extended to other cavity shape such as limacon cavity [26-29], quadruple cavity [20-21] et al.

## Conclusion

By utilizing a simple one-dimensional modulation in gain and loss, we have achieved single-mode laser operation with directional outputs in circular cavity. Compared with previous reports with modulation along Azimuthal direction, the one-dimensional modulation is much easier to be realized, e.g., by two beam interference. Most importantly, here the design also takes the advantage of grating effect of gain-loss modulation, which modifies the wave-vector of resonances and thus generates directional outputs. We believe our research will pave the applications of PT symmetric lasers.

## Acknowledgement

This work is supported by NSFC11204055, NSFC61222507, NSFC11374078, NCET-11-0809, KQCX2012080709143322, and KQCX20130627094615410, Shenzhen fundamental researches under the grant Nos. JCYJ20140417172417110, JCYJ20140417172417110, and JCYJ20140417172417096.

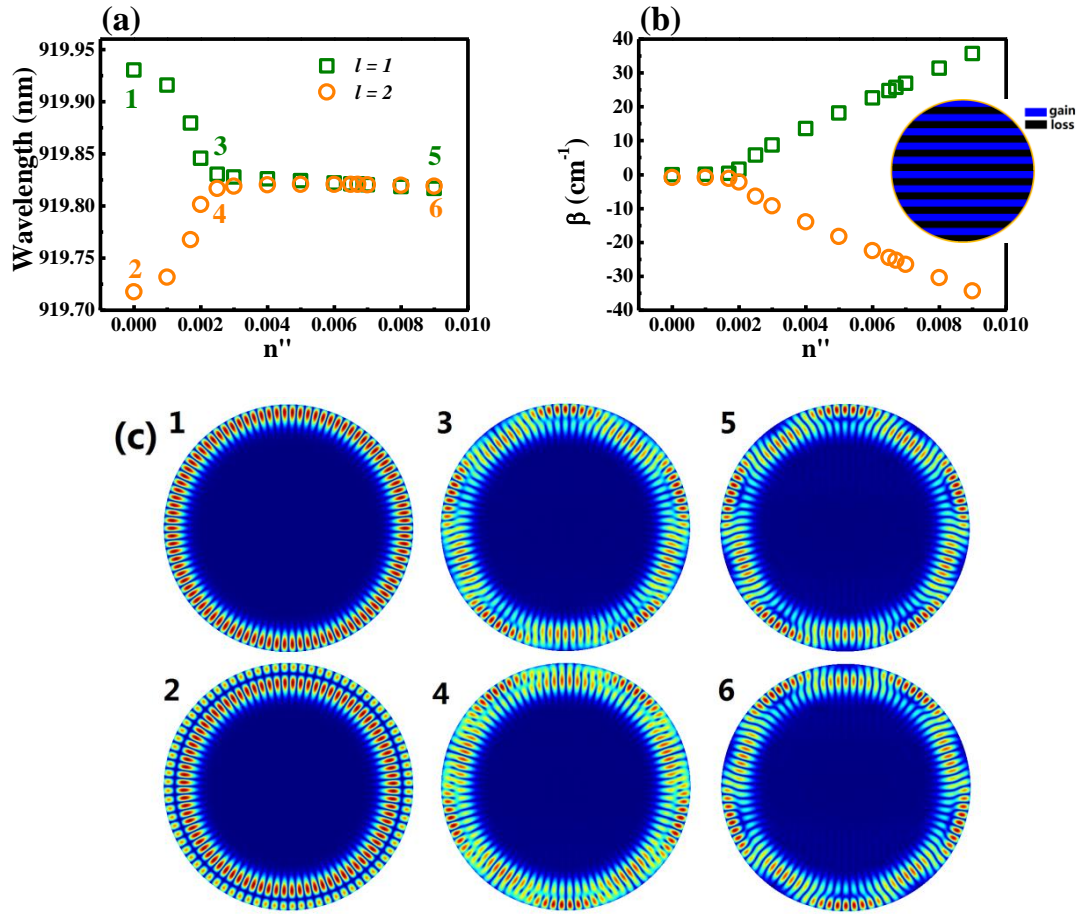

**Fig. 1:** (a) and (b) shows the PT symmetric behaviors of real and imaginary parts of lasing and absorptive modes. (c) shows the field patterns of modes 1-6 marked in (a). Inset in (b) is the schematic picture of the gain and loss distribution. The position around modes 3 and 4 is defined as the bifurcating point, where the imaginary parts of two modes starts to bifurcate. The field distributions of two modes are dramatically changed after the bifurcating point.

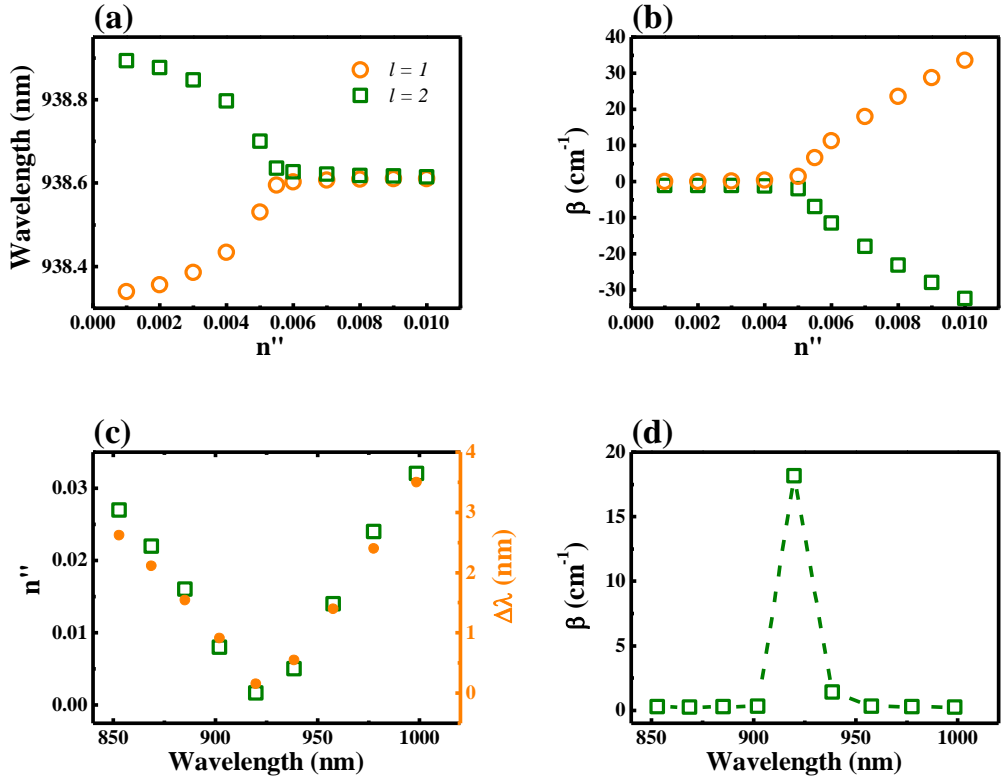

**Fig. 2:** (a) and (b) are the wavelength and b of mode at 938.6nm as a function of n". Due to the larger wavelength difference, bifurcating point of lasing and absorptive modes shifts to larger n". (c) The bifurcating point and wavelength detuning in passive cavity as a function of wavelength. (d) The β values of different resonances with n" = 0.005. It is easy to see that the mode at 919.8nm reaches the threshold first and always experiences the highest gain.

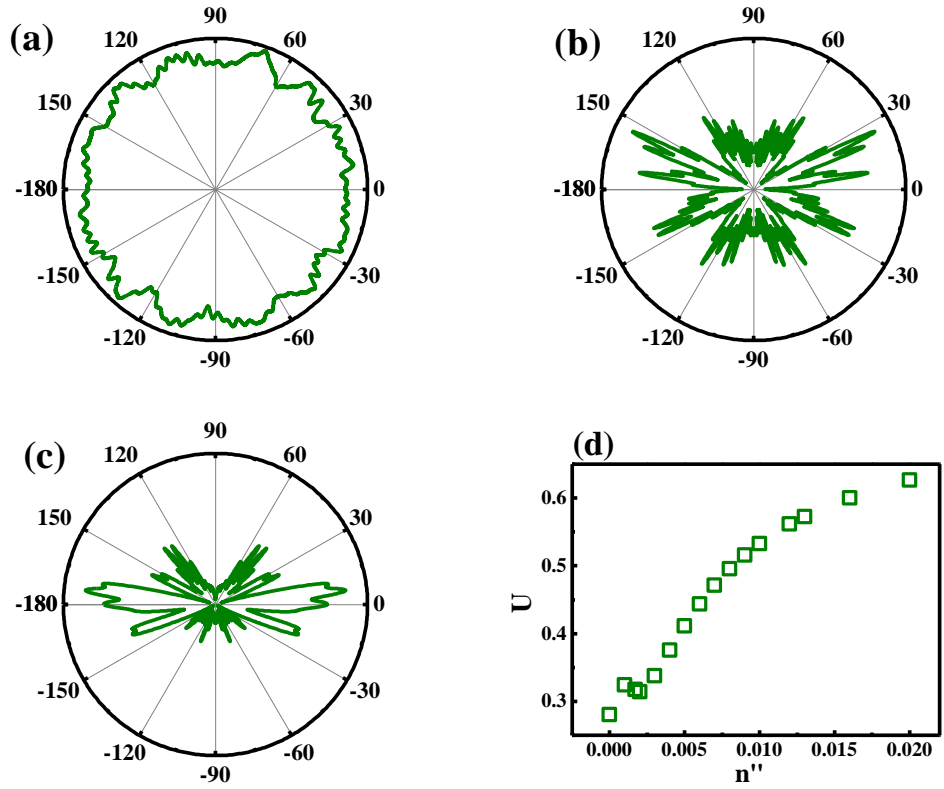

**Fig. 3:** Far field patterns of lasing mode with n" = 0.000 (a), n" = 0.0017 (b), and n" = 0.009 (c). (d) shows the dependence of directionality U on n". U is defined as $U = \left(\int_{150}^{200} I(\phi_{FF})d\phi_{FF} + \int_{-30}^{20} I(\phi_{FF})d\phi_{FF}\right) / \int_{-150}^{200} I(\phi_{FF})d\phi_{FF}$. Here $\phi_{FF}$ and $I(\phi_{FF})$ are the far field angle and its corresponding field intensity.

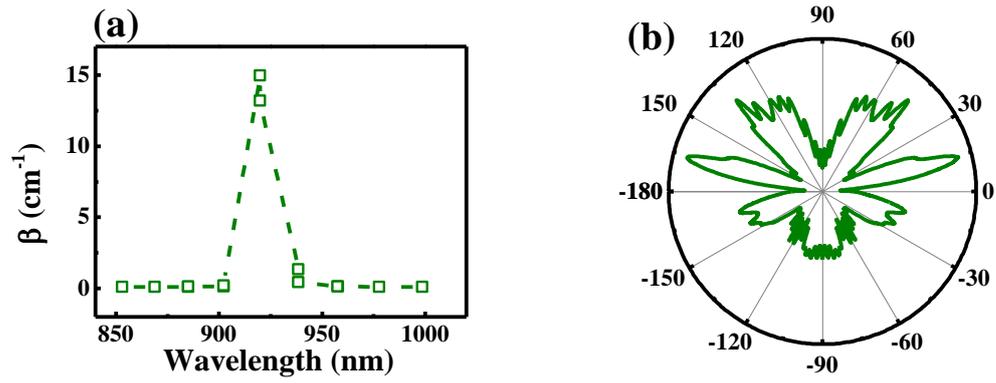

**Fig.4:** (a) The β values of the same whispering gallery modes as Fig. 2(d). (b) The far field pattern of the mode at 919.8 nm. Here the imaginary part of refractive index n" is 0.007.